\documentclass[article,onecolumn,superscriptaddress,nofootinbib,floatfix,onecolumn]{revtex4}

\usepackage{amsmath, latexsym, amssymb, hyperref, graphicx, color, multirow, makecell, diagbox}
\usepackage{tabularx}
\usepackage[T1]{fontenc}

\usepackage{dcolumn}
\usepackage{bm}
\usepackage{xspace}

\usepackage{float}
\usepackage{epsfig}
\usepackage{longtable}
\usepackage{rotating}
\usepackage{ulem}
\usepackage{amssymb}
\usepackage{amsmath}
\usepackage{txfonts}
\usepackage{upgreek}
\usepackage{mathtools}
\usepackage{tabularx}
\usepackage{booktabs}

\usepackage{soul} 

\usepackage[capitalize]{cleveref}
\definecolor{nicered}{rgb}{.7,.1,.1}
\definecolor{nicegreen}{rgb}{.1,.5,.1}
\definecolor{darkblue}{rgb}{0,0,.5}
\hypersetup{colorlinks, citecolor=nicegreen,linkcolor=nicered, urlcolor=darkblue}
\usepackage{multirow}
\usepackage{slashed}
\usepackage{verbatim}

\def\mean#1{\ensuremath{\left<#1\right>}}

\def\cO#1{{{\cal{O}}}\left(#1\right)}

\providecommand{\qqbar}{q\overline{q}}
\providecommand{\ccbar}{c\overline{c}}

\newcommand{\pp}{p-p}

\providecommand{\pPb}{p-Pb}

\providecommand{\PbPb}{Pb-Pb}
\newcommand{\ABgaga}{A\,B\,$\xrightarrow{\gaga}$ A\,$(\gaga)$\,B}

\newcommand{\epem}{e^+e^-}
\newcommand{\mumu}{\mu^+\mu^-}

\newcommand{\tata}{\mathcal{T}_0}

\newcommand{\gaga}{\gamma\gamma}

\newcommand{\etacOneS}{\mathrm{\eta_{c}(1S)}}
\newcommand{\etacTwoS}{\mathrm{\eta_{c}(2S)}}
\newcommand{\chicZeroP}{\mathrm{\chi_{c,0}(1P)}}
\newcommand{\chicTwoP}{\mathrm{\chi_{c,2}(1P)}}
\newcommand{\chicZero}{\mathrm{\chi_{c0}}}
\newcommand{\chicTwo}{\mathrm{\chi_{c2}}}

\providecommand{\mgg}{m_{\gamma\gamma}}

\newcommand{\sqrts}{\sqrt{s}}

\newcommand{\sqrtsnn}{\sqrt{s_{_\text{NN}}}}

\newcommand{\Lumi}{\mathcal{L}}

\newcommand{\LumiInt}{\mathcal{L}_{\mathrm{\tiny{int}}}}

\newcommand{\helaconia}{\textsc{HELAC-Onia}}
\newcommand{\madgraph}{\textsc{MadGraph5\_aMC@NLO}}

\def\mean#1{\ensuremath{\left<#1\right>}}

\usepackage{xspace}
\newcommand*{\eg}{e.g.,\@\xspace}
\newcommand*{\ie}{i.e.,\@\xspace}
\newcommand*{\cm}{c.m.\@\xspace}

\begin{document}

\title{Observing true tauonium via two-photon fusion at $\epem$ and hadron colliders}
 
\author{David~d'Enterria}\email{david.d'enterria@cern.ch}
\affiliation{CERN, EP Department, CH-1211 Geneva, Switzerland}
\author{Hua-Sheng~Shao}\email{huasheng.shao@lpthe.jussieu.fr}
\affiliation{Laboratoire de Physique Th\'eorique et Hautes Energies (LPTHE), UMR 7589,\\ Sorbonne Universit\'e et CNRS, 4 place Jussieu, 75252 Paris Cedex 05, France}

\begin{abstract}
\noindent
The feasibility of observing  true tauonium, the bound state of two tau leptons, $\tata\equiv(\tau^+\tau^-)_0$, via photon-photon collisions at $\epem$ colliders and at the LHC, is studied. The production cross sections of the process $\gaga\to\tata\to\gaga$ ---as well as those of all relevant backgrounds: spin-0 and 2 charmonium resonances decaying to diphotons, and light-by-light scattering--- are computed in the equivalent photon approximation for $\epem$ collisions at BES~III ($\sqrts = 3.8$~GeV), Belle~II ($\sqrts = 10.6$~GeV), and FCC-ee ($\sqrts = 91.2$~GeV), as well as for ultraperipheral \pp, \pPb, and \PbPb\ collisions at the LHC. Despite small $\tata$ production cross sections and a final state swamped by decays from overlapping pseudoscalar and tensor charmonium states ---the $\chicTwo$, $\etacTwoS$, and $\chicZero$ states have masses only 2.5, 84, and 139~MeV away, respectively, from the $\tata$ peak--- evidence and observation of the ground state of the heaviest leptonium appears feasible at Belle~II and FCC-ee, respectively, with in-situ high-precision measurements of the irreducible backgrounds.
\end{abstract}

\date{\today}

\maketitle

\nopagebreak

\section{Introduction}
Opposite-charge leptons ($\ell^\pm = e^\pm, \mu^\pm, \tau^\pm$) can form transient ``onium'' bound states under their quantum electrodynamics (QED) interaction. Like for the hydrogen atom, the various states of such exotic atoms feature a rich spectroscopic structure arising from the relative spin orientation of their two spin-$1/2$ leptonic constituents. The leptonium ground state (with principal quantum number $n=1$) has two states with total angular momentum $J=0$~and~1 known as para- and ortho-leptonium, respectively. On the one hand, spin-singlet para-leptonium states $1^1\mathrm{S}_0$ (using the spectroscopic $n^{2S+1}L_J$ notation, with total spin $S=0,1$, and orbital angular momentum $L=0,1,...\equiv \mathrm{S},\mathrm{P},...$) have leptonic constituents with antiparallel spins, they carry $J^{PC}=0^{-+}$ quantum numbers (with charge conjugation $C$, and parity $P$), and they decay preferentially into two photons.
On the other hand, triplet ortho-leptonium ($1^3\mathrm{S}_1$) states are composed of leptons with parallel spins, they feature $J^{PC}=1^{--}$ quantum numbers, and they decay into $3\gamma$ or, if kinematically accessible, into lighter $\ell^+\ell^-$ or quark-antiquark ($\qqbar$) final states. The most well-known leptonium system is positronium, discovered in 1951~\cite{Deutsch:1951zza}, whose spectroscopy has been thoroughly studied as a means to provide stringent tests of QED~\cite{Karshenboim:2005iy}, as well as in searches for violations of the discrete $CPT$ symmetries either singly or in various combinations~\cite{Bernreuther:1988tt,Yamazaki:2009hp}. The muonic counterpart of positronium, called true\footnote{The \textit{true} adjective is added to avoid any confusion with states composed of an electron plus a muon ($e^\pm\mu^\mp$), observed in 1960~\cite{Hughes:1960zz}, or a tau ($e^\pm\tau^\mp$).} muonium or dimuonium~\cite{Malenfant:1987tm}, has never been observed, nor the heaviest leptonium state, true tauonium or ditauonium $\mathcal{T}\equiv(\tau^+\tau^-)$. 
This work focuses on this latter system, barely studied since it was first suggested in~\cite{Moffat:1975uw,Avilez:1977ai,Avilez:1978sa}, with a mass of $m_{_{\tata}} = 2m_\tau + E_\text{bind} = 3553.696\pm 0.240$~MeV and binding energy of $ E_\text{bind}  = -\alpha^2 m_\tau/(4n^2) + \mathcal{O}(\alpha^4) = -23.7$~keV for $n=1$, using $m_\tau = 1776.86\pm0.12$~MeV and $\alpha= 1/137.036$~\cite{Zyla:2020zbs}. Recently~\cite{dEnterria:2022alo}, the diphoton decay width of para-ditauonium  $(n {^1}\mathrm{S}_0)$ has been derived including QED corrections up to next-to-next-to-leading-order accuracy,  finding
\begin{equation}
\Gamma_{\gaga}(\tata) = \frac{\alpha^5 \, m_\tau}{2\,n^3} \left(1 + 3.2725 (\alpha/\pi) + 97.12 (\alpha/\pi)^2\right)= 0.018533~\text{eV (for the \;}n=1\,\text{ state})\,.
\label{eq:gaga_tata_width}
\end{equation}
This decay width is about eight times larger than that of the tau lepton itself, $\Gamma_\text{tot}({\tau})= 0.00227$~eV~\cite{Zyla:2020zbs}, and, correspondingly, the $\tata$ diphoton lifetime\footnote{Natural units, $\hslash=c=1$, are used throughout the paper.} ($\tauup = 1/\Gamma \approx 36.0$~fs), is eight times smaller than that of the free $\tau$ lepton ($\tauup = 290.3$~fs). Para-ditauonium can thus be really produced as a $(\tau^+\tau^-)_0$ bound state before any of its constituent leptons decays weakly. Its total width is $\Gamma_\text{tot}(\tata) \approx \Gamma_{\gaga}(\tata) + 2 \Gamma_\text{tot}({\tau}) = 0.02384$~eV, with branching fractions $\mathcal{B}_{\gaga,\tau} = \Gamma_{\gaga,\tau}/\Gamma_\text{tot} \approx 80\%$ and 20\% for the diphoton and weak decays, respectively, neglecting few percent Dalitz decays contributions~\cite{dEnterria:2022alo}. This is at variance with ortho-ditauonium for which its three dominant decay channels\footnote{The 3-photon channel is much smaller: $\Gamma_{3\gamma}(\mathcal{T}_1) = 2(\pi^2 - 9)\,\alpha^6 m_\tau/(9\pi\,n^3) \approx \alpha^6 m_\tau/(16.3\,n^3)$.}---with widths $\Gamma_{\epem,\,\mu^+\mu^-}(\mathcal{T}_1)= \alpha^5 m_\tau/(6\,n^3)$ for the two leptonic modes (\ie $\mathcal{B}_{\epem,\,\mu^+\mu^-}(\mathcal{T}_1) \approx 21\%$ each), and
$\Gamma_{\qqbar}(\mathcal{T}_1) 
\approx 2\,\Gamma_{\epem,\,\mu^+\mu^-}$ for the hadronic mode (\ie $\mathcal{B}_{\qqbar}(\mathcal{T}_1) \approx 42\%$ for all light quarks inclusive)--- compete with the $\mathcal{B}_{\tau}(\mathcal{T}_1) \approx15\%$ single-tau weak decay branching fraction.\\

Since the tau lepton is 3500 and 17 times more massive, respectively, than the electron and muon, the ditauonium Bohr radius $a_0 = 2/(m_\tau\alpha)= 30.4$~fm is the smallest of all leptonium systems, 
and its associated minimum ``photon ionization'' energy (Rydberg constant), $R_\infty = \alpha/(4\pi a_0) = 3.76$~keV, is the largest. Namely, $\tata$ is the most strongly bound of all leptonia. Compared to precision studies of other exotic atoms, the investigation of ditauonium properties can thereby provide new tests of QED and of $CPT$ symmetries at high masses or, equivalently, small distances. First, the hyperfine structure and decay rates of ditauonium are influenced by QED (and QCD) vacuum polarization effects in the far time-like region, larger than those affecting lighter bound states such as dimuonium~\cite{Jentschura:1997tv}. 
Second, ditauonium features enhanced sensitivity to any physics beyond the standard model (BSM) that is suppressed by powers of $\cO{m_{\ell}/\Lambda_\text{BSM}}$ or affected by uncertainties from hadronic effects, as is the case for, \eg\ positronium or muonic-hydrogen states, respectively. The comparison of positronium, dimuonium, and ditauonium decays can thereby provide complementary information of any potential BSM, \eg\ lepton-flavor-violation effects, observed using the corresponding ``open'' leptons.

\begin{figure}[htpb!]
\centering
\includegraphics[width=0.95\textwidth]{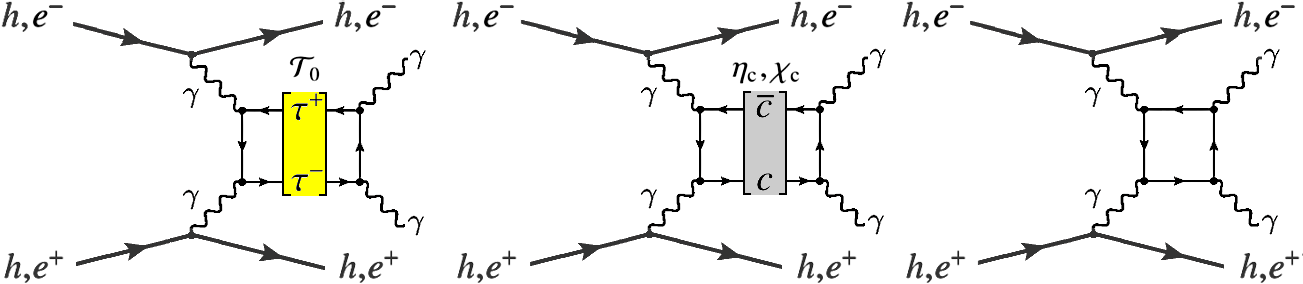}
\caption{Diagrams for two-photon production of para-ditauonium (left), $C$-even $\ccbar$ resonances (center), and LbL continuum (right), in hadron and $\epem$ collisions.}
\label{fig:gaga_tata_backgds}
\end{figure}

Three different production modes of the so-far unobserved dimuonium and ditauonium states have been considered at colliders: (i) $s$-channel production, or $t$-channel associated with $\gamma$ emission, in $\epem$ collisions~\cite{Brodsky:2009gx,Perl:1992xt}, (ii) photon-photon fusion in ultraperipheral collisions (UPCs) with heavy ions~\cite{Ginzburg:1998df,Baur:2001jj,Francener:2021wzx}, and (iii) very rare decays of heavier mesons produced in proton-proton (\pp) collisions~\cite{Fael:2018ktm,CidVidal:2019qub}. On the one hand, in $s$-channel $\epem$ annihilation only ortho-leptonium states with $J^{PC}=1^{--}$ 
can be produced and, given the narrow widths of all leptonia states, observing their resonant production is challenging due to losses from initial-state radiation and the need to monochromatize the beams to reduce their energy spread~\cite{Kirkby:1996qt,Bogomyagkov:2017uul,Telnov:2020rxp}. On the other hand, in collisions of two real photons $C$-even para-leptonium resonances 
can in principle be produced following the Landau--Yang theorem~\cite{Landau:1948kw,Yang:1950rg}. The production of $C=-1$ states can also proceed through the fusion of three (or, increasingly suppressed, five, seven, \dots) photons, as well as in the scattering of a \textit{virtual} plus a real photon~\cite{Schuler:1997yw}, but the corresponding cross sections are suppressed by at least a factor of $\alpha$~\cite{Ginzburg:1998df}. For the ditauonium case, simple estimates of the cross sections for the production of ortho-states in $\epem$ collisions at BEPC-II, as well as of para-states in \PbPb\ UPCs at the LHC, have been given in~\cite{Malik:2008pn} and~\cite{Baur:2001jj}, respectively. Additionally, a few $B$-hadron events decaying into ditauonium, $B \to K^{(*)} + \mathcal{T}$ with $\mathcal{O}(10^{-13})$ probability, are expected at the LHC with the total \pp\ integrated luminosities ($\LumiInt = 3$~ab$^{-1}$)~\cite{Fael:2018ktm}. However, none of the works above considered any actual experimental measurement of ditauonium, accounting for detector effects (acceptances, efficiencies, resolutions), physical backgrounds, nor associated uncertainties.\\

This study assesses for the first time the experimental feasibility of observing the production of true tauonium at present and future $\epem$ and hadron colliders. We consider the process of $\gaga$-fusion production of para-$\tata$, decaying into a pair of photons, shown in Fig.~\ref{fig:gaga_tata_backgds} (left).
 [Alternative production mechanisms and detection possibilities are covered in an upcoming work~\cite{DdEHSS}]. Diphoton backgrounds from $\gaga$ collisions leading to $C$-even charmonium resonances (Fig.~\ref{fig:gaga_tata_backgds}, center), and to the light-by-light (LbL) continuum (Fig.~\ref{fig:gaga_tata_backgds}, right)~\cite{dEnterria:2013zqi} are also computed. Four charmonium resonances have diphoton peaks in the vicinity of the $\tata$ signal: two pseudoscalar $\eta_c(1S,2S)$, one scalar $\chicZeroP$, and one tensor $\chicTwoP$ states, with their relevant properties listed in Table~\ref{tab:input}.

\begin{table}[htpb!]
\centering
\caption{$J^{PC}$ quantum numbers, mass $m_X$, total width $\Gamma_\text{tot}$, diphoton partial width $\Gamma_{\gaga}$, and $\mathcal{B}_{\gaga}$ branching fraction of para-ditauonium ($\tata$), and nearby pseudoscalar $\eta_c(1S,2S)$, scalar $\chicZeroP$, and tensor $\chicTwoP$ charmonium states~\cite{Zyla:2020zbs}.} 
\label{tab:input}
\begin{tabular}{lccccc}\toprule
Resonance &  $J^{PC}$ & $m_X$ (MeV) & $\Gamma_\text{tot}$ (MeV) & $\Gamma_{\gaga}$ (MeV) & $\mathcal{B}_{\gaga}$\\\midrule
$\tata$ & $0^{-+}$ & \;\;\texttt{$3553.696 \pm 0.240$}\;\; & \texttt{$2.28\cdot 10^{-8}$} &  \texttt{$1.83\cdot 10^{-8}$} & $\sim$80\% \\
$\etacOneS$ & $0^{-+}$ & \texttt{$2983.9 \pm 0.5$} & \texttt{$32.0 \pm 0.7$} & \texttt{$(5.06 \pm 0.34)\cdot 10^{-3}$} & $(0.0158 \pm 0.0011)\%$ \\
$\etacTwoS$ & $0^{-+}$ & \texttt{$3637.5 \pm 1.1$} & \texttt{$11.3 \pm 3.1$} & \texttt{$(2.15 \pm 1.47)\cdot 10^{-3}$} & $(0.019\pm 0.013)\%$ \\
$\chicZero$ & $0^{++}$ & \texttt{$3414.71  \pm 0.30$} & \texttt{$10.8 \pm 0.6$} & \texttt{$(2.203 \pm 0.097)\cdot 10^{-3}$} & $(0.0204 \pm 0.0009)\%$ \\ 
$\chicTwo$ & $2^{++}$ & \texttt{$3556.17 \pm 0.07$} & \texttt{$1.97\pm 0.09$} & \texttt{$(5.614 \pm 0.197)\cdot 10^{-4}$} & $(0.0285 \pm 0.0010)\%$ \\
\bottomrule
\end{tabular}
\end{table}

\section{Theoretical calculations}
\begin{figure}[htbp!]
\centering
\includegraphics[width=0.49\textwidth]{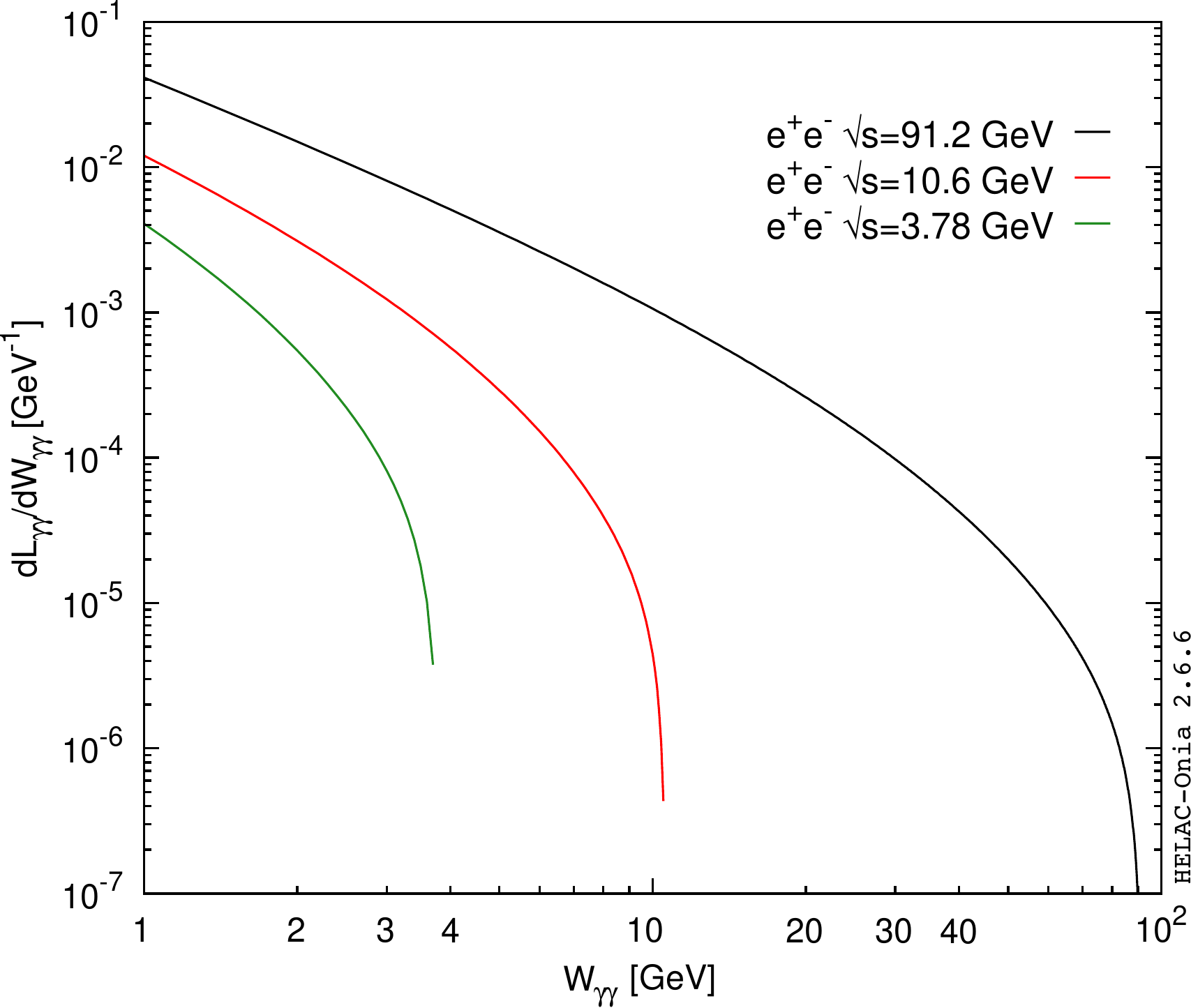}
\includegraphics[width=0.49\textwidth]{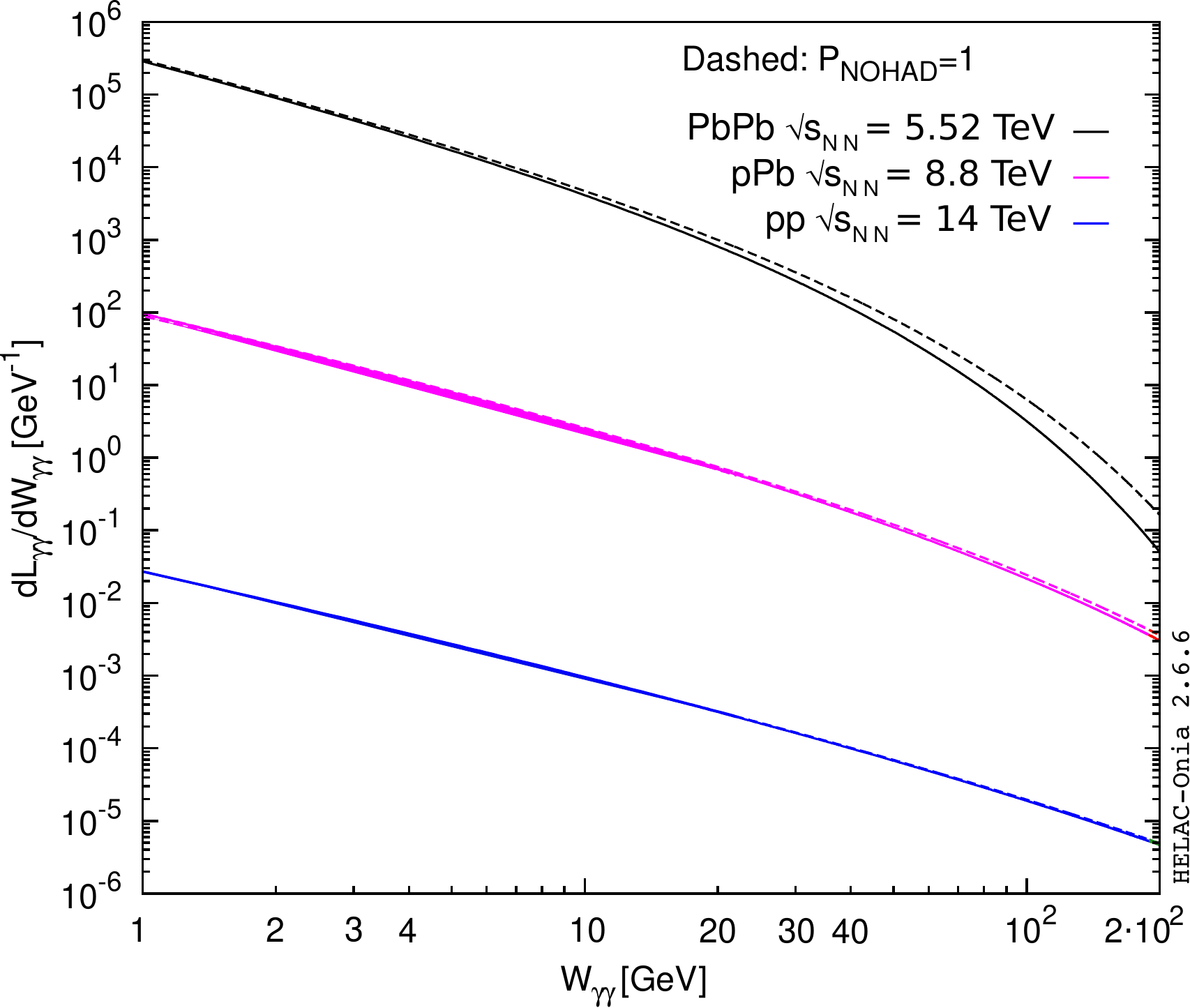}
\caption{Effective photon-photon luminosities $\mathrm{d}\Lumi_{\gaga}/\mathrm{d}W_{\gaga}$ as a function of $W_{\gaga}$ for the various $\epem$ (left) and LHC ultraperipheral (right) collisions considered here. The dashed curves in the right plot show the luminosities with nuclear overlap allowed.
\label{fig:lumi_gaga}}
\end{figure}

Cross sections for the signal and background processes shown in Fig.~\ref{fig:gaga_tata_backgds} are computed in the equivalent photon approximation (EPA)~\cite{Budnev:1974de} through a convolution of the elementary $\gaga\to\,X$ cross sections with the corresponding photon fluxes of the colliding beam particles. The exclusive production cross section of a $C$-even resonance $X$ through $\gaga$ fusion in a collision of charged particles $a\,b$ is given by 
\begin{equation}
\sigma(a\,b \to a\,b+X) = 4\pi^2 (2J+1)\frac{\Gamma_{\gaga}(X)}{m_X^2} 
    \left. \frac{\mathrm{d}{\Lumi}^{(a\,b)}_{\gaga}}{\mathrm{d}W_{\gaga}} \right|_{W_{\gaga}=m_X},
\label{eq:sigma_AA_X}
\end{equation}
where $W_{\gaga}$ is the $\gaga$ center-of-mass (\cm) energy, $m_X$ the mass of the resonance, and $\Gamma_{\gaga}(X)$ 
its two-photon width. The factor $\frac{\mathrm{d}{\Lumi}^{(a\,b)}_{\gaga}}{\mathrm{d}W_{\gaga}}\big|_{W_{\gaga}=m_X}$ is the value of the effective two-photon luminosity function at the resonance mass, determined from the convolution of the incoming photon EPA fluxes. For $\epem$ beams, the $\gamma$ flux is estimated with the Weizs\"acker-Williams approximation~\cite{Kniehl:1996we} (also cf. Eq.~(3) of~\cite{Flore:2020jau}), with the maximum virtuality set to $Q^2_\text{max} = 1$~GeV$^2$ as we focus on quasireal EPA scatterings, without the need to tag the transversely scattered $e^\pm$ in $\gaga$ collisions at high virtualities. For ion beams with charge number $Z$ and Lorentz boost $\gamma_\text{L}$, the photon number density at impact parameter $b$, derived from its corresponding electric dipole form factor, is $N_{\gamma/Z}(E_\gamma,b)=\frac{Z^2\alpha}{\pi^2}\frac{\xi^2}{b^2}[K_1^2(\xi)+\frac{1}{\gamma_\text{L}^2}K_0^2(\xi)]$, where $E_\gamma$ is the energy of the photon, $\xi=E_\gamma b/\gamma_\text{L}$, and $K_i$'s are modified Bessel functions~\cite{Baltz:2007kq}. The same expressions are applicable to proton beams using $Z=1$. At variance with the $\epem$ case, the effective $\gaga$ luminosity in UPCs with hadrons cannot be factorized as a direct convolution of the product of the photon fluxes of the two beams, due to the presence of a nonzero probability of hadronic interactions that break the exclusivity requirement. For an UPC \ABgaga\ at nucleon-nucleon \cm\ energy $\sqrtsnn$ of hadronic charges $Z_{1,2}$ with radii $R_{A,B}$, the $\gaga$ luminosity function reads
\begin{eqnarray}
\frac{\mathrm{d}{\Lumi}^\mathrm{(AB)}_{\gaga}}{\mathrm{d}W_{\gaga}}&=&\frac{2W_{\gaga}}{s_{_\text{NN}}}\int{\frac{\mathrm{d}E_{\gamma_1}}{E_{\gamma_1}}\frac{\mathrm{d}E_{\gamma_2}}{E_{\gamma_2}}\delta\left(\frac{W_{\gaga}^2}{s_{_\text{NN}}}-\frac{4E_{\gamma_1}E_{\gamma_2}}{s_{_\text{NN}}}\right)\frac{\mathrm{d}^2N^\mathrm{(AB)}_{\gamma_1/Z_1,\gamma_2/Z_2}}{\mathrm{d}E_{\gamma_1}\mathrm{d}E_{\gamma_2}}},
\end{eqnarray}
where the two-photon differential yield is
\begin{eqnarray}
\frac{\mathrm{d}^2N^\mathrm{(AB)}_{\gamma_1/Z_1,\gamma_2/Z_2}}{\mathrm{d}E_{\gamma_1}\mathrm{d}E_{\gamma_2}}&=&\int{\mathrm{d}^2\textbf{b}_1\mathrm{d}^2\textbf{b}_2\, P_{_\text{ no\,had}}\left(\left|\textbf{b}_1-\textbf{b}_2\right|\right)\,N_{\gamma_1/Z_1}(E_{\gamma_1},b_1)N_{\gamma_2/Z_2}(E_{\gamma_2},b_2)\,\theta(b_1-R_\mathrm{A})\theta(b_2-R_\mathrm{B})}\,,
\end{eqnarray}
and the probability to have no hadronic interaction at $b$ for nucleus-nucleus, proton-nucleus, and \pp\ collisions is given by~\cite{Klein:2016yzr}: $P_{_\text{ no\,had}}\left(b\right) = e^{-\sigma_{_\text{NN}}T_\text{AB}(b)}$, $e^{-\sigma_{_\text{NN}}T_\text{A}(b)}$, and $\big|1-\Upgamma(s_{_\text{NN}},b)\big|^2$, respectively;
where $T_\mathrm{A}(b)$ and $T_\mathrm{AB}(b)$ are the nuclear thickness and overlap functions, $\sigma_{_\text{NN}}$ is the inelastic NN scattering cross section parametrized as a function of $\sqrtsnn$~\cite{dEnterria:2020dwq}, and $\Upgamma(s_{_\text{NN}},b)$ is the Fourier transform of the \pp\ elastic scattering amplitude modeled by an exponential function $\Upgamma(s_{_\text{NN}},b)\approx e^{-b^2/(2b_0)}$ with $b_0=19.8$~GeV$^{-2}$~\cite{Frankfurt:2006jp}. Alternative prescriptions for the nonoverlap UPC condition, discussed in~\cite{HSS_DdE}, yield similar signal and background cross sections. Figure~\ref{fig:lumi_gaga} shows the effective $\mathrm{d}\Lumi_{\gaga}/\mathrm{d}W_{\gaga}$ luminosities for the different $\epem$ (left) and hadronic (right) colliding systems considered here. In the right plot, the dashed curves show the luminosities without the nonoverlap nuclear condition, which are mostly relevant for \PbPb\ collisions and start to be increasingly visible above $W_{\gaga}\approx 10$~GeV. 
Although photon-fusion processes in \PbPb\ (\pPb) UPCs benefit from  the lack of pileup collisions and from a huge $Z^2$ ($Z$) charge enhancement factor compared to the \pp\ case, the orders-of-magnitude larger proton beam $\LumiInt$, and the availability of forward proton detectors to tag such collisions, eventually compensate for this difference above $W_{\gaga}\approx$~100~GeV~\cite{Bruce:2018yzs}.\\

The \helaconia\ 2.6.6 Monte Carlo (MC) code~\cite{Shao:2012iz,Shao:2015vga} complemented with the EPA photon setup discussed above is used to compute the cross sections and generate the $\tata$ signal and charmonia backgrounds events~\cite{HL}. A modified version of the $\eta_c(1S)$ particle with $3553.7$-MeV mass and 
$1.83\cdot10^{-8}$-MeV width is implemented to emulate the $\tata$ signal. The $\tata$ finite width (and corresponding lifetime) is accounted for by reshuffling the momentum of the resonance according to its associated Breit--Wigner (B--W) distribution~\cite{Frixione:2019fxg}. Spin-correlated diphoton decays of the tensor $\chicTwo$ meson are implemented following the formula derived in~\cite{Shao:2012fs,Shao:2014yfa}. The loop-induced LbL background is simulated with \madgraph\ v2.6.6~\cite{Alwall:2014hca,Hirschi:2015iia}, with the virtual box contributions computed at leading order. Table~\ref{tab:xsecs} lists (middle columns) the computed cross sections times $\mathcal{B}_{\gaga}$ for signal and backgrounds in the different colliding systems. The relative production cross sections are given by the proportions $\etacOneS:\chicTwoP:\chicZeroP:\etacTwoS:\tata \approx 100:50:30:25:1$, basically driven by their different $\Gamma^2_{\gaga}/(\Gamma_\text{tot}\cdot m_X^2)$ ratios as per the product of Eq.~(\ref{eq:sigma_AA_X}) times $\mathcal{B}_{\gaga}$. The cross sections for para-ditauonium are factors $\times$25--100 smaller than for the rest of resonances, mostly because of the narrow diphoton width of this state (Table~\ref{tab:input}) that leads to a correspondingly smaller photon-fusion production probability. The exponentially decreasing LbL continuum below the ditauonium peak is about $\approx$100 larger than the signal peak, but can be largely removed with appropriate kinematic criteria and/or constrained in mass sidebands free of any resonance peak. We note that the production cross sections for excited $n^1\mathrm{S}_0$ para-ditauonioum states, densely (few keV) spaced above the $\tata$ mass, can be derived from that of the $1^1\mathrm{S}_0$ ground state via $\sigma(n^1\mathrm{S}_0) = \sigma(1^1\mathrm{S}_0)/n^3$, following Eq.~(\ref{eq:gaga_tata_width}). Such higher orbital para-ditauonium states have $n^3$-suppressed diphoton widths, and thereby smaller $\gaga$-fusion production cross sections and longer lifetimes that compete with their single-tau weak decays. Since these excited states will add only a few percent contributions to the diphoton yields within the Gaussian-smeared $\tata$ ground-state peak, they are neglected hereafter.

\setlength{\tabcolsep}{6pt}
\begin{table}[htpb!]
\centering
\caption{Photon-fusion production cross sections $\sigma\times\mathcal{B}_{\gaga}$ for para-ditauonium signal and backgrounds ($C$-even charmonium states, and LbL scattering over $m_{\gaga}\in (m_{_{\tata}} \pm 100$~MeV), and $|\eta_\gamma|<5$) decaying to diphotons, at various $\epem$ facilities and in UPCs at the LHC. The last column lists the total produced $\tata$ and dominant irreducible $\chicTwo$ yields 
for the integrated luminosities quoted at each collider (those for the LHC correspond to LHCb). Uncertainties (not quoted) are around $\pm10\%$ (except for $\etacTwoS$, see text).
\label{tab:xsecs}}
\begin{tabular}{l|cccccc|cc} \hline
Colliding system, \cm\ energy, $\LumiInt$, exp.\ & \multicolumn{6}{c|}{$\sigma\times\mathcal{B}_{\gaga}$} & 
\multicolumn{2}{c}{$N\times\mathcal{B}_{\gaga}$}\\
& \;\;$\etacOneS$ & $\etacTwoS$ & $\chicZeroP$ &  $\chicTwoP$ & LbL & $\tata$ & \;\;$\tata$\;\; & \;\;$\chicTwoP$\;\; \\ \hline
$\epem$ at 3.78~GeV, 20 fb$^{-1}$, BES~III & 120 fb & 3.6 ab & 15 ab & 13 ab & 30 ab   &  0.25 ab &  -- & -- \\
$\epem$ at 10.6~GeV, 50 ab$^{-1}$, Belle~II & 1.7 fb & 0.35 fb & 0.52 fb & 0.77 fb & 1.7 fb  & 0.015 fb &  750  & 38\,500\\
$\epem$ at 91.2~GeV, 50 ab$^{-1}$, FCC-ee \;& 11 fb & 2.8 fb & 3.9 fb & 6.0 fb & 12 fb & 0.11 fb &  5\,600 & $3\cdot10^{5}$ \\\hline
\pp\ at 14 TeV, 300~fb$^{-1}$, LHC & 7.9 fb & 2.0 fb & 2.8 fb & 4.3 fb & 6.3 fb  & 0.08 fb &  24  & 1290 \\
\pPb\ at 8.8 TeV, 0.6~pb$^{-1}$, LHC & 25 pb & 6.3 pb & 8.7 pb & 13 pb & 21 pb  & 0.25 pb &  0.15 & 8 \\
\PbPb\ at 5.5 TeV, 2~nb$^{-1}$, LHC & 61 nb & 15 nb & 21 nb & 31 nb & 62 nb  & 0.59 nb &  1.2 & 62\\
\hline
\end{tabular}
\end{table}

The uncertainties of the theoretical cross sections quoted in Table~\ref{tab:xsecs} can be estimated from the ingredients of Eq.~(\ref{eq:sigma_AA_X}). The relative uncertainties of the $\Gamma_{\gaga}$ and $\Gamma_\text{tot}$ widths of all resonances propagate into their final $\sigma\times\mathcal{B}_{\gaga}$  cross section scaled by a factor of two and linearly, respectively. They are negligible for $\tata$ and, added in quadrature, amount to relative uncertainties in the 8\%--14\% range (except for $\etacTwoS$, which is of $\pm140\%$ due to its currently badly known diphoton width). 
Uncertainties related to the $\gaga$ effective luminosities and the nonoverlap condition in the case of UPCs~\cite{HSS_DdE} are of the same order but affect all resonance cross sections in a fully correlated manner. In any case, all background cross sections uncertainties can be significantly reduced with in-situ measurements of all the charmonium resonances while, or prior to, performing the $\tata$ signal extraction, as described below.

\section{Experimental feasibility}
The hierarchy of effective $\gaga$ luminosities shown in Fig.~\ref{fig:lumi_gaga} indicates that the higher the \cm\ energy and the beam charges, the larger the expected $\tata$ cross sections (Table~\ref{tab:xsecs}). At BES~III, the cross sections are in the subattobarn range because the \cm\ energies of interest are only reached in the very suppressed tail of the colliding $\gamma$ fluxes, as $\sqrts = 3.78$~GeV is not very far from the $W_{\gaga} = m_{\tata}$ threshold for the production of the resonance. The Super-KEKB (Belle~II)~\cite{Kou:2018nap} and FCC-ee at the Z pole (91~GeV)~\cite{Abada:2019zxq} appear as the most interesting facilities in terms of $\tata$ production yields, thanks to the huge $\LumiInt = 50$~ab$^{-1}$ values expected at both machines. At the LHC, although the $\tata$ production cross sections in UPCs with ions are larger by orders of magnitude (up to the nb range) compared to $\epem$ and \pp\ systems, the possibility to reconstruct its relatively soft decay photons, $E_\gamma\approx m_{\tata}/2 =$~1.5--2~GeV, with good enough acceptance and energy resolution appears only feasible at the LHCb experiment~\cite{LHCb:2018roe}. Unfortunately, this experiment features  $\LumiInt$ values (quoted in Table~\ref{tab:xsecs}) about ten times smaller than those of ATLAS/CMS (and of ALICE for \pPb\ and \PbPb) leading to small final numbers of visible signal events. In addition, in hadronic collisions one has to deal with the extra production of the $C$-even charmonium resonances in central exclusive (gluon-induced) processes with much larger cross sections than the photon-fusion ones~\cite{Harland-Lang:2010ajr}, which increase the backgrounds (although imposing low final-state $\gaga$ acoplanarities largely reduces them~\cite{dEnterria:2013zqi}).\\

\begin{figure}[htpb!]
\centering
\includegraphics[width=0.531\textwidth]{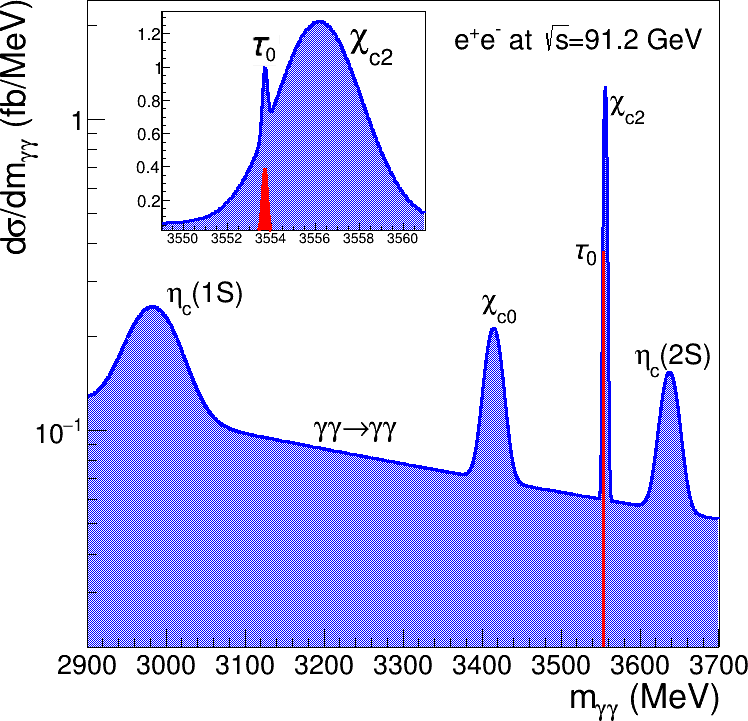}
\includegraphics[width=0.464\textwidth]{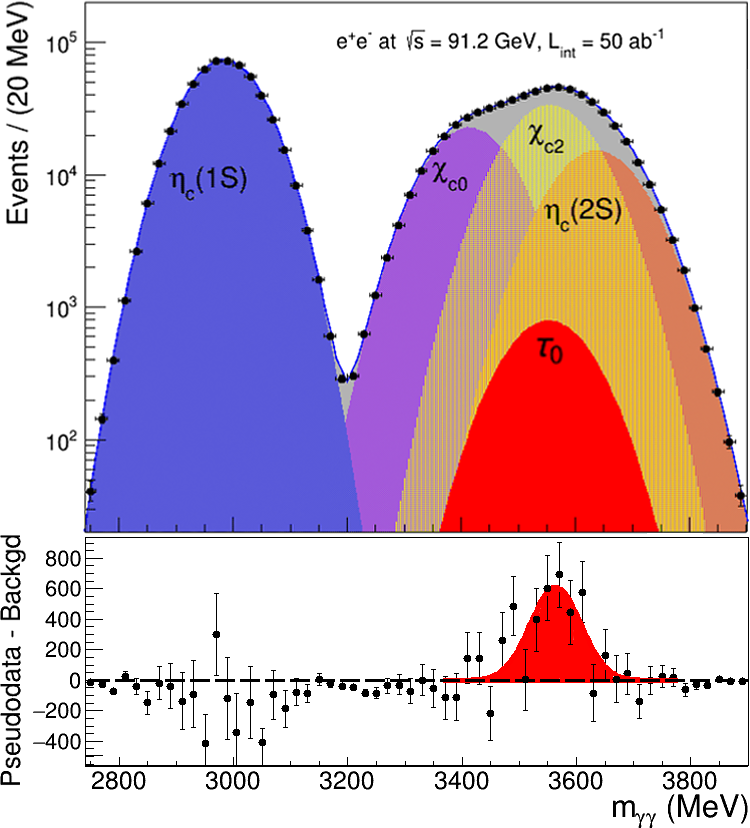}
\caption{Diphoton invariant mass distribution from photon-fusion processes in $\epem$ collisions expected at FCC-ee over $\mgg \approx 2.8$--3.8~GeV, shown as theoretical cross sections (left, with the $\tata$ width arbitrarily set to 0.1~MeV to make it visible) and as number of counts accounting for experimental resolution (right, with the LbL continuum subtracted). The bottom panel of the right plot, shows the pseudodata minus the background-only fit.
\label{fig:Minv_gaga_tata_genlevel}}
\end{figure}

Hereafter, we thereby focus on the feasibility of an experimental measurement of ditauonium at Belle~II and FCC-ee (91~GeV), with $\sim$750 and $\sim$5600 para-$\tata$ events produced, respectively. The theoretical diphoton mass distributions are shown in Fig.~\ref{fig:Minv_gaga_tata_genlevel} (left) for the FCC-ee case. The experimental signature of interest is that of a final state with two photons exclusively produced, \ie\ without any other activity in the event, emitted back-to-back in azimuth, and with an invariant mass peaked at $m_{\tata} = 3.5537$~GeV. A trigger can be easily set up that selects online events with two photons with such generic properties with $\sim$100\% efficiency, while suppressing processes such as, \eg\ $\epem \to \gaga$ annihilation, which feature diphoton invariant mass peaking at the actual \cm\ energy $\sqrts\gg m_{\tata}$. One million events are generated with \helaconia\ and \madgraph\ for the para-$\tata$ signal and for all individual backgrounds listed in Table~\ref{tab:xsecs} for (asymmetric) $4+7$~GeV and (symmetric) $45.6+45.6$~GeV $\epem$ collisions at Belle~II and FCC-ee, respectively. For both experimental setups, a polar angle photon acceptance of $10^\circ < \theta_\gamma < 170^\circ$, a $\sim$2\% diphoton mass resolution, and $\sim$100\% $\gamma$ reconstruction efficiencies, consistent with Belle-II performances~\cite{Kou:2018nap}, are assumed. Such an angular acceptance keeps, respectively, about 95\% and 63\% (resp., 82\% and 60\%) of the resonances and LbL yields at Belle~II (resp., FCC-ee). The Gaussian smearing of the reconstructed $\gamma$ energy broadens the narrow B--W diphoton peaks to $\sim$70~MeV experimental widths 
as shown in Fig.~\ref{fig:Minv_gaga_tata_genlevel} (right). We note that the cross section of $\gaga\to\etacOneS\to\gaga$ can be measured with outstanding precision and accuracy thanks to 85\,000 and 500\,000 counts, isolated from any other neighboring resonance, at Belle~II and FCC-ee, respectively. This resonance can thus be used as a ``standard candle'', first, to accurately monitor \textit{in situ} the photon energy calibration and resolution, as well as to precisely determine all diphoton reconstruction efficiencies and associated uncertainties. The measurement of $\sigma(\etacOneS)\times\mathcal{B}_{\gaga}$ will, in addition, allow an accurate control and validation of all ingredients of the theoretical photon-fusion calculations, Eq.~(\ref{eq:sigma_AA_X}), which are common to the $\tata$ signal and all backgrounds.\\

The data analysis strategy follows three basic steps: (i) reconstructing the $\gaga$ invariant mass distribution of all events passing a set of multivariate analysis (MVA) cuts defined so as to reduce backgrounds while keeping the largest $\tata$ signal possible, 
(ii) fitting the overall $m_{\gaga}$ distribution to a model combining all resonant peaks with positions and widths fixed to their most precise values, 
and subtracting the underlying exponential-like LbL continuum, 
and (iii) performing a profile likelihood ratio analysis of two different fits, which assume the presence or absence of the signal in the (pseudo)data samples, in order to extract the final $\tata$ yield and its statistical significance. An MVA is first performed on the simulated samples exploiting twelve different kinematic variables of the final-state single $\gamma$ and $\gaga$ pairs. We use the TMVA package~\cite{Hocker:2007ht} to train and test boosted decision-tree (BDT) classifiers to provide statistical discrimination between the ditauonium signal and the backgrounds, and maximize the signal significance. Cutting on an appropriate value of the BDT response, which exploits the different single and pair photon kinematics of LbL and $\tata$, allows the reduction of the LbL continuum by a factor of $\sim20$ with no signal loss. The remaining nonresonant LbL distribution can be reproduced with an exponential fit. 
Unfortunately, the BDT is unable to find any discriminating power (other than the mass, which will be used in the final fit) against the $\ccbar$ resonances that overlap with the signal peak. This is not surprising for the (pseudo)scalar charmonia, as they share the same kinematic properties of the para-ditauonium state, but it was not \textit{a priori} expected for the tensor $\chicTwo$ meson. The fact remains that although the $\chicTwo$ decay photon angular distribution is partially different from that of the pure scalar $\tata$ state, its yields are $\sim$50 times larger than those of the signal, and the best significance found by the BDT analysis corresponds to a working point that keeps the maximum number of signal counts without any effective background reduction.\\

In the final step of the analysis, a fit is performed of the simulated diphoton invariant mass distribution over $m_{\gaga} \approx 2.8$--3.8~GeV with two models: the default one that combines the expected $\tata$ signal plus all backgrounds, and the null-hypothesis that assumes no para-ditauonium signal to be present. In the fit, all resonances are fixed at their nominal masses 
with yields normalizations fixed to their theoretical predictions, with statistical uncertainties corresponding to $\LumiInt = 50$~ab$^{-1}$ at Belle~II and FCC-ee, and with systematic uncertainties assigned as explained next.\\

First, as aforementioned, the $\etacOneS$ state is of no concern for the $\tata$ extraction as it does not have any overlap with the signal for the expected photon energy resolution, and plays no actual role in the fit. Second, for the $\chicZero$ and $\etacTwoS$ mesons that partially overlap the $\tata$ signal, one can identify ranges of their diphoton lineshapes (\eg\ between $\mgg\approx 3200$--3400~MeV and 3750--3900~MeV, respectively), where both charmonium states can be measured virtually free from any contamination from other nearby resonances. This will allow for a first estimation of their signal contamination with $\mathcal{O}($1\%) systematic uncertainties. 
In addition, one can exploit the large $\gaga\to\chicZero, \etacTwoS$ samples available, amounting to $\sim$10--100 million events at Belle~II and FCC-ee, to measure alternative decays with $\mathcal{O}(100)$ times larger branching fractions than the diphoton one (such as, \eg\ 
the four-meson $\chicZero$~\cite{Belle:2007qae} and three-meson $\etacTwoS$~\cite{BaBar:2014asx,Belle:2018bry} channels). All such measurements can provide ultraprecise determinations of the $\chicZero$ and $\etacTwoS$ total and diphoton widths and, thereby, an accurate control of their corresponding contamination in the $\tata$ signal region. The third case is that of the largest $\chicTwo\to\gaga$ background. Since it almost perfectly overlaps with the $\tata$ signal mass region, an independent precise determination of its diphoton width (and associated $\gaga$ cross section) is mandatory prior to any attempt to extract the $\tata$ signal. To independently measure the $\Gamma_{\text{tot},\gaga}(\chicTwo)$ widths, one can exploit the very large event data samples produced in two-photon fusion but decaying into alternative \textit{charged}-particle final states, free of any $\tata$ contribution and with an accurate momentum resolution that allows for a reconstruction of its natural B--W shape with $\Gamma_\text{tot}(\chicTwo) \approx 1.97$~MeV. The $\chicTwo\to \pi^+\pi^-\pi^+\pi^-$ decay, with a branching fraction of 1\% (\ie\ 36 times larger than that of $\chicTwo\to\gaga$), provides a potential data sample of about 1.4 (11) million events at Belle~II (FCC-ee). A fit of the exclusive 4-charged-pion invariant mass distribution around $m_{\chicTwo}$ to the expected B--W shape for this resonance, would lead to an extraction of the $\chicTwo$ natural width at Belle~II (FCC-ee) with a statistical uncertainty about twenty (fifty) times smaller than the $\sim$9\%  value of the current LHCb state-of-the-art measurement in the $\chicTwo\to J\psi(\mu\mu)\mu\mu$ channel~\cite{LHCb:2017hzb}. 
By combining different $\chicTwo$ decay modes, it is not unreasonable to achieve a few per-mil precision in the lineshape of this charmonium resonance. For the purpose of this study, the final fit is carried out assigning to $\sigma(\chicTwo)\times\mathcal{B}_{\gaga}$ a relative systematic uncertainty of 0.3\% and 0.2\% at Belle~II and FCC-ee, respectively.\\

The statistical significance of the $\tata$ signal is derived 
from the likelihood ratio of the two fits: background-only imposing $N(\tata) = 0$, and the default signal-plus-background.
Statistical significances of about 3 and 5 standard deviations (std. dev.) are obtained at Belle-II and FCC-ee, respectively. The bottom panel of Fig.~\ref{fig:Minv_gaga_tata_genlevel} (right) shows the residual distribution of the pseudodata minus the null-hypothesis fit. More sophisticated statistical analyses could be considered ---\eg\ by carrying out the likelihood study in different categories of diphoton angular ranges, where the relative contributions of scalar $\tata$ and tensor $\chicTwo$ decays vary within a few tens of percent--- that would slightly increase the signal significance. Nonetheless, this first exploratory study indicates that the observation (evidence) of the production of para-ditauonium is feasible in $\gaga$ collisions at FCC-ee (Belle~II), respectively.\\ 

Before closing, it is worth mentioning that one could singularly observe the production of para-ditauonium 
by exploiting its narrow natural width (relatively large lifetime) that leads to a decay away from the $\epem$ interaction vertex. Indeed, whereas all charmonium resonances decay almost immediately after production, the $\sim$30-fs total lifetime of $\tata$ leads to an exponential tail of secondary decay vertices that could be used to uniquely identify the production of this exotic QED atom away from any background. At Belle~II, the produced para-$\tata$ is relatively slow 
($\mean{\beta\gamma} \approx 0.8$, including the asymmetric $\beta\gamma\approx 0.28$ beam boost) and has a mean production vertex of $\mean{L_\text{vtx}}\approx 10~\mu$m fairly indistinguishable from the primary $\epem$ collision one. At FCC-ee Z-pole energies, the Lorentz boost is larger ($\beta\gamma\approx 3$) and results in a mean tauonium decay length of $\mean{L_\text{vtx}}\approx 30~\mu$m, with counts expected in the tail up to $\sim$0.1~mm. On the one hand, the diphoton vertex pointing capabilities are coarse, in the 1-cm range for LHC-type calorimeters, unless one uses converted photons and/or high-precision time-of-flight (separating mm distances requires few ps timing resolution)~\cite{Dudar:2021ybu}. On the other hand, one could alternatively exploit the $\tata\to\epem\gamma$ and $\tata\to\mumu\gamma$ Dalitz decays, with combined branching fraction of $\mathcal{O}(3\%)$~\cite{dEnterria:2022alo}, leading to $\mathcal{O}(100)$ or $\mathcal{O}(20)$ signal counts at FCC-ee and Belle-II, respectively. For such decays, one can use the much more accurate charged-particle secondary vertex capabilities of the dielectron and/or dimuon system to uniquely identify 
the $\tata$ state as a displaced resonance. Exploiting such an alternative possibility requires a dedicated study that goes beyond this first work.

\section{Summary}
We have presented the first feasibility study to produce and observe the bound state of two tau leptons (true tauonium or ditauonium), the heaviest and most compact purely leptonic ``atomic'' system. Ditauonium remains experimentally unobserved to date, and can be exploited for novel bound-state QED tests sensitive to physics beyond the standard model that does not impact its lighter siblings (positronium, and dimuonium). 
Cross sections for the photon-fusion production of the para-ditauonium ground state $\tata$ decaying into two photons, $\gaga\to\tata\to\gaga$, have been calculated for current and future $\epem$ and hadron colliders. The largest cross sections (in the nb range) are achieved in ultraperipheral collisions of heavy ions at the LHC, but the much higher integrated luminosities in $\epem$ collisions at Belle~II and FCC-ee favor both facilities in terms of expected yields (750 and 5600 counts, respectively). A realistic study of the experimental measurement of $\tata$ production on top of the large diphoton backgrounds from light-by-light scattering and overlapping $C$-even charmonium resonances ($\chicZero, \chicTwo$, and $\etacTwoS$) has been performed. The very large expected data samples of $\chicZero, \chicTwo$, and $\etacTwoS$ mesons produced via photon-photon collisions will allow for a control of their contamination in the signal region at the subpercent level. A multivariate analysis combined with a multi-Gaussian fit indicates that evidence (3 std.\ dev.) and discovery (5 std.\ dev.) of ditauonium appear feasible at Belle~II and FCC-ee, respectively.
The new photon-photon collision framework implemented here in the \helaconia\ and \madgraph\ packages~\cite{HL} can be also exploited for studies of $C$-even charmonium production at the LHC~\cite{Chapon:2020heu}, as well as for searches for axionlike particles decaying into two photons~\cite{dEnterria:2021ljz} at present and future $\epem$ and hadron colliders.\\

\paragraph*{Acknowledgments.---} D.~d'E. thanks useful discussions with R.~Perez-Ramos and L.~Gouskos on different aspects of this work. Support from the European Union's Horizon 2020 research and innovation program (grant agreement No.824093, STRONG-2020, EU Virtual Access ``NLOAccess''), the French ANR (grant ANR-20-CE31-0015, ``PrecisOnium''), and the CNRS IEA (grant No.205210, ``GlueGraph"), are acknowledged.




\bibliographystyle{myutphys}
\bibliography{reference}

\end{document}